\documentclass[12pt]{article}

\makeatletter
\usepackage{theorem}
\def\ifundefined#1{\expandafter\ifx\csname#1\endcsname\relax}
     \theorembodyfont{\slshape}
        \newtheorem{thm}{Theorem}
     \newtheorem{prop}[thm]{Proposition}

     \theorembodyfont{\upshape}
     \newtheorem{defn}[thm]{Definition}
     \newtheorem{const}[thm]{Construction}
     
     \newtheorem{example}[thm]{Example}

\ifundefined{proof}

\newcommand{\proofname}{Proof}
\fi
\usepackage{amsfonts}
\newcommand{\keywords}[1]{\begingroup \def \protect {\noexpand \protect 
\noexpand }\xdef \@thefnmark { }\endgroup \@footnotetext{{\em Keywords and 
phrases.\/} #1}}
\newcommand{\AMSMSC}[2]{\begingroup \def \protect 
{\noexpand \protect \noexpand }\xdef \@thefnmark { }\endgroup 
\@footnotetext{{1991 \it Mathematical Subject Classification.\/} Primary: 
#1; Secondary: #2.}}

\def\p@enumi{}
\makeatother

\newcommand{\comment}[1]{}

\newcommand{\algebra}[1]{\ensuremath{{\frak #1}}}
\newcommand{\Heisen}[1]{\ensuremath{{{\mathbb H}^{#1}}}}
\newcommand{\hHeisen}[1]{\ensuremath{{{ \widehat{\mathbb H}^{#1}}}}}

\newcommand{\Space}[2]{\ensuremath{ {{\mathbb #1}^{#2}} }}

\newcommand{\FSpace}[2]{{\ensuremath{ #1_{#2} }}}

\ifundefined{qed}
    \DeclareMathSymbol{\qed}{0}{AMSa}{"03}
\fi

\newcommand{\eqref}[1]{\textup{Eq.~(\ref{#1})}}

\newcommand{\F}{{\mathcal F}}
\newcommand{\R}{{\mathbb R}}
\newcommand{\Z}{{\mathbb Z}}
\newcommand{\C}{{\mathbb C}}
\newcommand{\p}{\textsl{p}} 
\newcommand{\pp}{\textsl{P}}

\newcommand{\changeone}[1]{#1}
\title{Mixing Quantum and Classical Mechanics}

\author{Oleg V. Prezhdo\thanks{E-mail: \texttt{oleg@czar.cm.utexas.edu}}\\
\emph{Department of Chemistry}\\
\emph{and Biochemistry}\\
\emph{University of Texas at Austin}\\
\emph{Austin, Texas 78712, USA}
\and
Vladimir V. Kisil\thanks{Current address: Vakgroep Wiskundige Analyse,
                       Universiteit Gent,
                       Galglaan 2, B-9000,
                       Gent, BELGIE. 
E-mail: \texttt{vk@cage.rug.ac.be}} \\
\emph{Institute of Mathematics}\\
\emph{Economics and Mechanics}\\
\emph{Odessa State University}\\
\emph{ul. Petra Velikogo, 2,}\\
\emph{Odessa-57, 270057, UKRAINE}
}

\date{October 11, 1996}

\begin{document}
\maketitle
\setlength{\baselineskip}{7mm}

\begin{abstract}
\setlength{\baselineskip}{7mm}

Using a group theoretical approach we derive an equation of motion for
a mixed quantum-classical system.
The quantum-classical bracket entering the equation preserves the
Lie algebra structure of quantum and classical mechanics:
The bracket is antisymmetric and satisfies the Jacobi identity, and, 
therefore, leads to a natural description of interaction between quantum 
and classical degrees of freedom. We apply the formalism to coupled quantum 
and classical oscillators and show how various approximations, 
such as the mean-field and the multiconfiguration mean-field approaches, 
can be obtained from the quantum-classical equation of motion.

\vspace{3mm}

\end{abstract}

\section{Introduction}\label{sec1}

Many phenomena in nature are described by quantum mechanics at a
fundamental level
and with high precision.
Yet, there exist numerous situations where mixed quantum-classical
models are needed.  In some cases the phenomena are too complex to
allow for a fully quantum approach, in others a consistent quantum
theory is lacking.
Classical mechanics often provides a more suggestive description and
a clearer picture of physical events.  Applications of various
quantum-classical approaches range from biochemical and condensed
matter chemical reactions, where the large dimensionality of the
systems of interest requires approximations, to the evolution of the
universe and cosmology, where no theory of quantum
gravity has been established.

The issue of treating quantum and classical degrees of freedom within
the same formalism was recently discussed in a number of
publications~\cite{Maddox95,Anderson95,Boucher88}. The interest
was spurred by the cosmological problem of defining the backreaction
of quantum matter fields on the classical space-time background,
where classical variables should be independently correlated with
each individual quantum state.  The traditional approach fails
to satisfy the last requirement.
(For a fully quantum approach to cosmology see
reference~\cite{ChamConnes96b}.)
Somewhat earlier a similar situation
was encountered in chemical physics, where quantum-classical
trajectory methods were employed to model gas phase scattering
phenomena and chemical dynamics in
liquids~\cite{Diestler83,Aleksandrov81,%
Miller80,Tully76,Pechukas69a,Pechukas69b}.
It was noticed in these studies that asymptotically distinct quantum
evolutions should correlate with different classical trajectories.

The first relationship between quantum and classical variables
is due to Ehrenfest~\cite{Ehrenfest27}
who showed that the equation of motion for the average values of
quantum observables coincides with the corresponding classical
expression. (Surprisingly, the first mathematically rigorous treatment
on the subject was not carried out until 1974, see
reference~\cite{Hepp74}.)
Ehrenfest's result leads to the mean-field approach, where classical
dynamics is coupled to the evolution of the expectation values of
quantum variables~\cite{Mott31,Mittelman61,Delos72a,Delos72b}. The
mean-field equations of motion possess all of the properties of the
purely classical equations and are exact as far as the mean
values of quantum operators are concerned.
However, an expectation value does
not provide information of the outcome of an individual process.
The mean-field approach gives a satisfactory
description of the classical subsystem as long as changes within
the quantum part are fast compared to the characteristic
classical time-scale.
If classical trajectories depend strongly on a particular
realization of the quantum evolution, the mean field approximation is
inadequate.  The problem can be corrected, for instance by
introduction
of stochastic quantum hops between preferred basis states with
probabilities determined by the usual quantum-mechanical
rules~\cite{Tully71,Tully90,OPrezhdo96b}.

Similarity between the algebraic structures underlying quantum and
classical mechanics provides a consistent way of improving upon the
mean-field approximation, as explored in
references~\cite{Anderson95,Boucher88,Aleksandrov81}.
\comment{references~\cite{Anderson95,Anderson95a,
Salcedo95,Boucher88,Aleksandrov81}.}
In those studies the aim was to derive a quantum-classical bracket
that reduces to the quantum commutator and the Poisson bracket in
the purely quantum and classical cases.  In addition to the reduction
property the bracket should satisfy other criteria so as to
give physically meaningful pictures of quantum-classical
evolutions.  In particular, an antisymmetric bracket conserves
the total energy and a bracket satisfying the Jacobi identity ensures
that the Heisenberg uncertainty principle is not violated.

Recently, one of us (VVK) proposed~\cite{Kisil96a} a natural
mathematical
construction, which we name \p-mechanics,
enveloping classical and quantum mechanics.
Formulated within the framework of operator algebras,
the \p-mechanical equation of motion reduces to the appropriate
quantum or
classical equations under suitable representations of the algebra
of observables.  In this paper we extend the ideas of \p-mechanics
to incorporate mixed
quantum-classical descriptions.  In particular, we derive the
quantum-classical bracket and explicitly show that it satisfies the
properties common to quantum and classical mechanics.
\comment{To our knowledge, this is the most consistent
derivation to date.  It provides a clear understanding of the
interaction between quantum and classical variables and allows for
further generalizations.}
Using the technique described it is possible to construct families of
mixed
quantum-classical approaches, each having a specific set of
properties.

\changeone{
The format of this paper is as follows:  In Section~\ref{sec2}
we summarize \p-mechanics and introduce the essential mathematical
definitions.  In Section~\ref{sec3} we
construct the simplest \p-mechanical model that adopts two distinct
sets of variables associated with quantum and classical degrees of
freedom.
By taking the appropriate
representation we derive the quantum-classical bracket and show that
it is antisymmetric and obeys the Jacobi identity, that is, it
possesses the two major properties shared by the quantum and
classical brackets.  In Section~\ref{se:2oscilators}
we work out the case of coupled
classical and quantum harmonic oscillators.
\comment{as an example of application of the suggested formalism}
Finally, in Section~\ref{sec4} we discuss how various
approximations to the general quantum-classical
description can be obtained, including the mean-field and
the multiconfiguration mean-field approaches.}




\section{\pp-mechanics}\label{sec2}
\comment{In this section we summarize the unified approach to 
classical and
quantum mechanics. Starting
with an algebraic construction it allows to derive classical
and quantum mechanical laws of evolution by selecting a
representation
for the generalized equation of motion.}

\subsection{The Elements of \pp-mechanics}
We recall the constructions from
references~\cite{Kisil96a,Kisil94e} together with appropriate
modifications.
\begin{defn}\label{de:p-mech}\textrm
An operator algebra \algebra{P} gives a \emph{\p-mechanical
description}~\cite{Kisil96a} of a system if the following
conditions hold.
\begin{enumerate}
\item\label{it:representation} The set $\widehat{\algebra{P}}$ of all
irreducible representations $\pi_h$ of \algebra{P} is a disjoint
union of subsets
$\widehat{\algebra{P}}=\sqcup_{p\in P} \widehat{\algebra{P}}_p$
parameterized by the elements of a set $P$. The elements of the set
$P$
are associated with different values for the Planck constant.
We refer to this set as the
\emph{set of Planck constants}. If for
$p_0$ the set $\widehat{\algebra{P}}_{p_0}$ consists of only
commutative (and, therefore, one-dimensional) representations, then
$\widehat{\algebra{P}}_{p_0}$ gives a \emph{classical} description. If
$\widehat{\algebra{P}}_{p_0}=\{\pi_{p_0}\}$ consists of a single
non-commutative representation $\pi_{p_0}$, then
$\widehat{\algebra{P}}_{p_0}$ gives a purely \emph{quantum} model.
Sets
$\widehat{\algebra{P}}_p$ of other types provide \emph{mixed}
(quantum-classical) descriptions.

\item\label{it:topology} Let $\widehat{\algebra{P}}$ be equipped with
a
natural operator topology (for example, it may be the Jacobson
topology~\cite{Dixmier69}  or the *-bundle
topology~\cite{DaunHof68,Hofmann72}). Then $P$ has a natural
\emph{factor
topology} induced by the partition $\widehat{\algebra{P}}=\sqcup_{p\in
P}$.

\item\label{it:dynamics}(Dynamics) The algebra \algebra{P} is equipped
with the one-parameter semigroup of transformations
$G(t):\algebra{P}\rightarrow\algebra{P}$, $t\in\Space{R}{+}$. All sets
$\widehat{\algebra{P}}_p$, $p\in P$ are \emph{preserved} by $G(t)$.
Namely, for any $\pi\in \widehat{\algebra{P}}_p$ all new
representations
$\pi_t=\pi\circ G(t)$ again belong to $\widehat{\algebra{P}}_p$.
\item \label{it:correspond}(The Correspondence Principle) Let
$S:p\mapsto S(p)\in \algebra{P}_p$ be an operator-valued section
continuous in the *-bundle topology~\cite{DaunHof68,Hofmann72} over
$P$.
Then for any $t$, i.e., at any moment of time the image
$S_t(p)=G(t)S(p)$ is also
a section due to statement~\ref{it:dynamics}.  In the *-bundle
topology
the sections $S_t(p)$ are \emph{continuous for all} $t$.
\end{enumerate}
\end{defn}

Having listed these quite natural conditions, we do not yet know how
to construct
\p-mechanics.  Next, we describe an important particular case
of \emph{group quantization}~\cite{Kisil94e}. All components of
\p-mechanics (operator
algebra, partition of representations, topology) readily arise there.

\begin{const}\rm \emph{Group quantization} consists of the following
steps.
\begin{enumerate}

\item Let $\Omega=\{x_j\}, 1\leq j\leq N$ be a set of physical
variables defining the state of a classical system.  Classical
observables are
real-valued functions on the states.

The best known and the most important case is the set
$\{x_j=q_j,x_{j+n}=p_j\},\
1\leq j\leq n, N=2n$ of coordinates and momenta of classical particles
forming an $n$ degree of freedom system. The observables are real
valued functions on
\Space{R}{2n}. We will use this example throughout this Section.

\item We complete the set $\Omega$ with
additional variables ${x_j}, N<j\leq \bar{N}$, such that the new set
$\bar{\Omega}$
forms the smallest algebra, which contains $\Omega$ and is closed
under
the Poisson bracket
\begin{displaymath}
\{x_i,x_j\}\in \bar{\Omega},\ \mbox{ for all } x_i,x_j\in
\bar{\Omega}.
\end{displaymath}

In the above example we add the unit function $x_{2n+1}=1$.
The complete set contains $\bar{N}=2n+1$ elements satisfying the
famous relations
\begin{equation}\label{eq:poisson}
\{x_j,x_{j+n}\}=-\{x_{j+n},x_j\}=x_{2n+1}.
\end{equation}
All other Poisson brackets are zero.

\item We form an $\bar{N}$-dimensional Lie algebra $\algebra{p}$ with
the
frame $\{\widehat{x}_j\},\ 1\leq j \leq \bar{N}$ defined by the formal
mapping $\hat{}: x_j\mapsto\widehat{x}_j$. The commutators of the
frame
vectors are formally defined by the formula
\begin{equation}\label{eq:hat-poisson}
[\widehat{x}_i,\widehat{x}_j]=\widehat{\{x_i,x_j\}}.
\end{equation}
We extend the commutator onto the whole algebra by linearity.

For our example, $\algebra{p}$ is the Lie algebra corresponding to
the Heisenberg group (see the next Subsection for details).

\item We introduce the algebra $\algebra{P}$ of
convolutions induced by \algebra{p}. The convolution
operators are \emph{observables} in the group quantization, and by
analogy with the classical case they can be treated as functions of
$\widehat{x}_j$. \changeone{Particular representations of the
convolution algebra in spaces $\Space{L}{2}(S)$ give different
descriptions of a physical system. The family of all
one-dimensional representations of
\algebra{P} corresponds to \emph{classical} mechanics;
various noncommutative representations lead to \emph{quantum} and
\emph{quantum-classical}
descriptions with different \emph{Planck constants}.}

For our example the following possibilities exist.
\begin{enumerate}
\item $S =\Space{R}{n},\ \widehat{x}_j=X_j=M_{q_j},\
\widehat{x}_{j+n}=-i\hbar\partial/\partial q_j$,
the convolutions are represented by pseudo-differential operators
(PDO), and we
obtain the \emph{Dirac-Heisenberg-Schr\"odinger-Weyl quantization}
by PDO.

\item $S =\Space{R}{2n},\ \widehat{x}_j=X_j=M_{q_j},\
\widehat{x}_{j+n}=M_{p_j}$, the convolutions are represented by
(operators of multiplication by) functions,
and we obtain the classical description that we started from.
\end{enumerate}
\end{enumerate}
\end{const}

It is an empirical observation that \emph{the steps above lead to a
nilpotent Lie group, with the dual $\widehat{Z}$ of the center $Z$ of
the group
interpreted as the set of Planck constants}. Now we
illustrate this fact by a well-known example of quantization, and
later in Section~\ref{sec3} by constructing a quantum-classical model.

\subsection{The Heisenberg Group Generates Quantum and Classic
Mechanics}
In the previous Subsection we claimed that the $n$th order Heisenberg
group
$\Heisen{n}$ describes a set of quantum particles that constitute an
$n$-degree
of freedom system.  Here we show how this description is achieved.

$\Heisen{n}$ is generated by the $n$-dimensional
translation and multiplication operators $e^{ip\cdot D}$, $e^{iq\cdot
X}$,
$p,q \in \Space{R}{n}$ satisfying the Weyl commutator relations
\begin{eqnarray}
e^{2\pi ip\cdot D}e^{2\pi iq\cdot X} &=&
e^{2\pi ip\cdot q}e^{2\pi iq\cdot X}e^{2\pi ip\cdot D}.
\label{eq:Wcommutator}
\end{eqnarray}
An element of the Heisenberg group $g\in\Heisen{n}$
is defined by $2n+1$ real numbers
$(p,q,s)$, $p,q\in\Space{R}{n} $, $s\in\Space{R}{}$.
The composition of two elements $g$ and $g'$ is given by
\begin{displaymath}
(p,q,s)(p',q',s')=(p+p',q+q',s+s'+\frac{1}{2}(pq'-p'q)).
\end{displaymath}
$D_j$, $X_j$, and $I$ form a $2n+1$ dimensional basis of the
Heisenberg algebra $\algebra{h}^n$ with a one-dimensional center
$\Z = \{sI; s\in\R\}$.
Since all second and higher order commutators of the basis
elements vanish,
$\Heisen{n}$ and $\algebra{h}^n$ are step two nilpotent
Lie group and algebra respectively.

The unitary irreducible representations of the Heisenberg group are
classified by
the Stone-von Neumann theorem~\cite{Folland89}.
They are parameterized by a real number
$h$, the character of the one-dimensional center $\Space{Z}{}$.
A non-zero $h$ gives non-commutative unitary representations acting on
the Hilbert
space $L^2(\R^n)$
\begin{eqnarray}
\rho_{h\ne 0}(p,q,s) &=& e^{2\pi i (p\cdot hD + q\cdot X + s\cdot
hI)}.
\label{eq:rho_h}
\end{eqnarray}
The $n$ components of $X$ and $hD$ are the usual quantum mechanical
position
$X_j$ (multiplication by $x_j$) and momentum $hD_j$ ($-i\hbar$ times
differentiation
with respect to $x_j$) operators characterized by the Heisenberg
commutator relation
\begin{eqnarray}
[hD_j,X_k] &=& -i h \delta_{jk} I.
\label{eq:Hcommutator}
\end{eqnarray}
In the limit of zero $h$ the center ${\Z}$ of the Heisenberg group
vanishes, and
$\Heisen{n}$ becomes isomorphic to $\R^{2n}$. The irreducible
representations
of the latter are homomorphisms from $\R^{2n}$ into the circle group
acting on ${\C}$
\begin{eqnarray}
\rho_{h=0}(p,q) &=& e^{2\pi i (pk + qx)}.
\label{eq:rho_0}
\end{eqnarray}
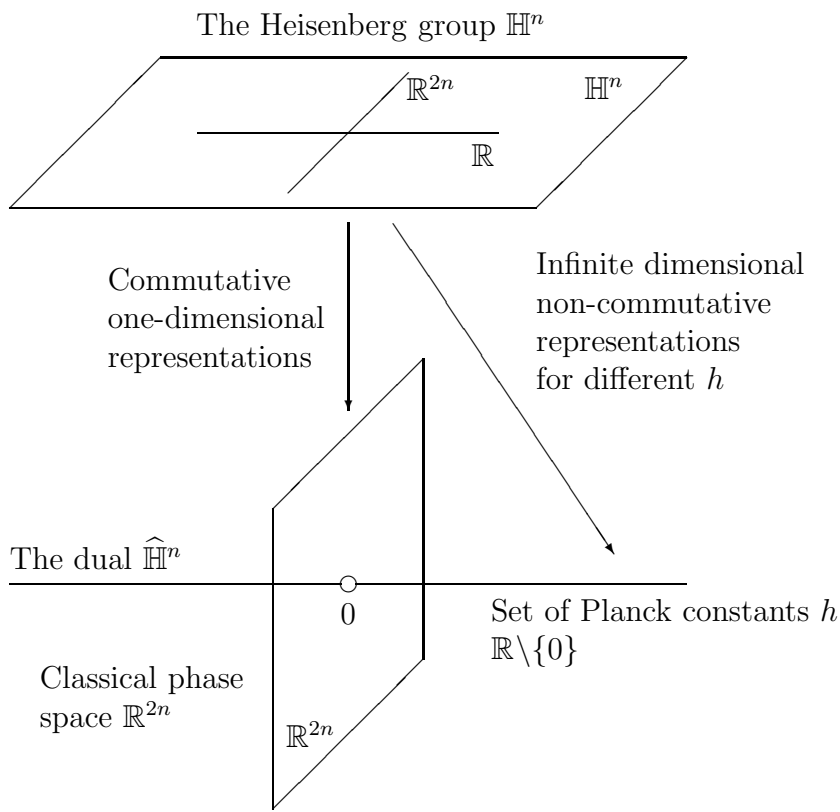
\begin{figure}[t]
\begin{center}
\special{em:linewidth 0.4pt}
\unitlength 1.00mm
\linethickness{0.4pt}
\begin{picture}(97.00,109.00)
\put(7.00,85.00){\line(1,1){20.00}}
\put(27.00,105.00){\line(1,0){70.00}}
\put(97.00,105.00){\line(-1,-1){20.00}}
\put(77.00,85.00){\line(-1,0){70.00}}
\put(62.00,65.00){\line(-1,-1){20.00}}
\put(42.00,45.00){\line(0,-1){40.00}}
\put(42.00,5.00){\line(1,1){20.00}}
\put(62.00,25.00){\line(0,1){40.00}}
\put(53.00,35.00){\line(1,0){44.00}}
\put(51.00,35.00){\line(-1,0){44.00}}
\put(52.00,35.00){\circle{2.00}}
\put(52.00,83.00){\vector(0,-1){25.00}}
\put(58.00,83.00){\vector(2,-3){29.33}}
\put(55.00,109.00){\makebox(0,0)[cc]{{The Heisenberg group \Heisen{n}}}}
\put(52.00,31.00){\makebox(0,0)[cc]{{$0$}}}
\put(71.00,33.00){\makebox(0,0)[lt]
{\parbox{50\unitlength}{Set of Planck constants $h$\\$\Space{R}{}\backslash \{0\}$}}}
\put(41.00,24.00){\makebox(0,0)[rt]
{\parbox{30\unitlength}{Classical phase\\ space $\Space{R}{2n}$}}}
\put(61.00,70.00){\makebox(0,0)[rc]
{\parbox{41\unitlength}{Commutative\\ one-dimensional\\ representations}}}
\put(77.00,70.00){\makebox(0,0)[lc]
{\parbox{50\unitlength}{Infinite dimensional\\ non-commutative \\ representations\\ for different $h$}}}
\put(44.00,87.00){\line(1,1){16.00}}
\put(32.00,95.00){\line(1,0){40.00}}
\put(86.00,101.00){\makebox(0,0)[cc]{\Heisen{n}}}
\put(70.00,92.00){\makebox(0,0)[cc]{\Space{R}{}}}
\put(63.00,101.00){\makebox(0,0)[cc]{\Space{R}{2n}}}
\put(7.00,37.00){\makebox(0,0)[lb]{{The dual $\hHeisen{n}$}}}
\put(47.00,15.00){\makebox(0,0)[cc]{{\Space{R}{2n}}}}
\end{picture}
\end{center}
\caption{The Heisenberg group and its dual.}\label{fi:dual}
\end{figure}
The dual $\hHeisen{n}$ as a set is equal to
$\{\Space{R}{}\setminus 0 \}\cup \Space{R}{2n}$ (see
Figure~\ref{fi:dual}). It has the natural topology coinciding
on $\{\Space{R}{}\setminus 0 \}$ with the Euclidean topology.  Any
sequence of representations $\{\rho_{h_j}\}$, $h_j\rightarrow 0$,
$h_j\neq 0$ is dense in whole $\Space{R}{2n}$. The last property is
fundamental for the \emph{correspondence principle}.

The unitary representations of $\Heisen{n}$ can be extended to the
convolution algebra
$L^1(\Heisen{n})$. Namely, if $A\in L^1(\Heisen{n})$, then
it defines a convolution on the Heisenberg group
\begin{displaymath}
A\cdot b(g) = \int_{\Heisen{n}} A(g') b(g\star g') dg'.
\end{displaymath}
The representation $\rho_h$ maps the convolution to the operator
\begin{eqnarray}
\rho_h(A) &=&
\int_{\Heisen{n}} A(g) \rho_h(g) dg \nonumber \\
&=& \int\int\int A(p,q,s) \rho_h(p,q,s) dp dq ds .
\label{eq:convolution}
\end{eqnarray}
The \p-mechanical equation of motion (see~\cite{Kisil96a} for
details) for an element $A(g)$ ($g\equiv\{p,q\})$
of the convolution algebra is defined by
\begin{eqnarray}
\frac{\partial A(g)}{\partial t } &=& 2\pi i [H,A](g)
\label{eq:eom}
\end{eqnarray}
with
\begin{eqnarray}
[H,A](g) &=& \int_{\Heisen{n}} [H(g')A(g'\ast g) - A(g')H(g'\ast g)]
dg' \nonumber,
\end{eqnarray}
where $H(g)$ is the Hamiltonian.
The non-commutative unitary representations of \eqref{eq:rho_h}
reduce this equation to the Heisenberg equation of motion for
operators acting on the Hilbert space $L^2(\R^n)$.  Under the
commutative
representations of \eqref{eq:rho_0} the \p-mechanical equation of
motion becomes
the Hamilton equation for functions on the phase space $\R^{2n}$.

We consider the last statement in more detail
by means of the pseudo-differential calculus directly related
to these group theoretical developments and the problem of
quantization.
The non-commutative unitary representations of
the Heisenberg group allow one to define integral operators
corresponding to
functions on phase space.  Given a function $\sigma(k,x)$ on
${\R}^{2n}$
one obtains the operator $\sigma(D,X)$ on $L^2(\R^n)$ by the formula
\begin{eqnarray}
\sigma(hD,X) &=&
\int_{\Heisen{n}} \F^{-1}[\sigma](g) \rho_{h\ne 0}(g) dg \nonumber \\
&=& \int\int \F^{-1}[\sigma](p,q) e^{2\pi i(p\cdot hD + q\cdot X)} dp
dq,
\label{eq:T}
\end{eqnarray}
where $\F^{-1}[\bullet]$ denotes the inverse Fourier transform.
The trivial integration over $s$
has been carried out.
The action of the operator $\sigma(D,X)$ on a function $f(x)\in
L^2(\R^n)$
follows from the definition of $hD$ and $X$ [see \eqref{eq:rho_h}
and the related paragraph], and is given by
\begin{eqnarray}
\sigma(hD,X) f(x) &=& \int\int \F^{-1}[\sigma](p,q) e^{\pi ihpq + 2\pi
iqx}
f(x+hp) dp dq \nonumber \\
&=& h^{-n} \int\int \F^{-1}[\sigma](\frac{y-x}{h},q) e^{\pi iq (x +
y)} f(y) dy dq
\nonumber \\
&=& h^{-n} \int\int \sigma(k,\frac{x+y}{2}) e^{2\pi i(x-y)k/h} f(y) dy
dk
\label{eq:Tf}
\end{eqnarray}
or
\begin{eqnarray}
\sigma(hD,X) f(x) &=& \int K_{\sigma}(x,y) f(y) dy,  \nonumber \\
K_{\sigma}(x,y) &=&
h^{-n} \int \sigma(k,\frac{x+y}{2}) e^{2\pi i(x-y)k/h} dk,
\label{eq:Kf}
\end{eqnarray}
where $K_{\sigma}$ is the kernel of the integral operator
$\sigma(hD,X)$.
In the language of the pseudo-differential calculus the function
$\sigma(k,x)$ is called
the symbol of the operator $\sigma(hD,X)$. If instead of $\rho_{h\ne
0}$ one uses
a commutative representation $\rho_{h=0}$, the
transformation of~\eqref{eq:T}
reduces to identity and we recover the \emph{classical observable}
$\sigma(k,x)$.
Eqs.~(\ref{eq:T})--(\ref{eq:Kf}) are known as the Weyl correspondence
principle.

The symbol $\sigma\sharp_h\tau$(k,x) of the product of two operators
$\sigma\sharp_h\tau(hD,X) = \sigma(hD,X)\cdot \tau(hD,X)$ can be
obtained
by application of a non-commutative representation to the convolution
on the
Heisenberg
group [see \eqref{eq:convolution}] or directly from the Weyl rule. It
is given
in terms of the symbols of individual operators as
\begin{eqnarray}
\sigma\sharp_h\tau(k,x) &=& \bigg(\frac{2}{h}\bigg)^{2n}
\int\int\int\int
\sigma(\zeta,u)\tau(\eta,v)
e^{4\pi i[(x-u)(k-\eta)-(x-v)(k-\zeta)]/h} du dv d\eta d\zeta.
\label{eq:product}
\end{eqnarray}
It follows from the discussion above that the non-commutative
representations of the Heisenberg group transform
the \p-mechanical equation of motion (\ref{eq:eom}) into the equation
for operators on $L^2(\R^n)$
\begin{eqnarray}
\frac{\partial}{\partial t}A(hD,X) &=& \frac{2\pi i}{h}
[H,A]_{\sharp_h}(hD,X)
\label{eq:heisenberg}
\end{eqnarray}
where $[H,A]_{\sharp_h}\equiv [H\sharp_h A - A\sharp_h H]$,
the operation of taking the product of two symbols $\sharp_h$ is
defined by
\eqref{eq:product}, and the operators $A(hD,X)$ and
$[H,A]_{\sharp_h}(hD,X)$ are recovered from their symbols $A(k,x)$ and
$[H,A]_{\sharp_h}(k,x)$ by the application of the Weyl transform
Eqs.~(\ref{eq:T})--(\ref{eq:Kf}).  This is the quantum-mechanical law
of
motion in the Heisenberg form.

In order to obtain the corresponding classical expression it is useful
to cast
the product rule of~\eqref{eq:product} in the form
of an asymptotic expansion in powers of $h$.  The integration
over $\eta$ and $\zeta$ and the change of variables $(u-x)/h
\rightarrow u$,
$(v-x)/h \rightarrow v$ converts \eqref{eq:product} to
\begin{eqnarray} \sigma\sharp_h\tau(k,x) &=& h^{-2n}
\int\int \F^{-1}_1[\sigma](v,x+uh) \F_1[\tau](u,x+vh) e^{4\pi i(v-u)k}
du dv,
\nonumber
\end{eqnarray}
where $\F_1$ and $\F_1^{-1}$ denote the Fourier transform and its
inverse
with respect to the first variable only. Expanding $\sigma$ and $\tau$
in the
second variable around $x$ and applying the Fourier inversion formula
to each term in the Taylor series we obtain
\begin{eqnarray}
\sigma\sharp_h\tau(k,x) &=& \sum_{\alpha+\beta\leq\gamma}
\frac{(i\pi h)^{\alpha+\beta}(-1)^{\alpha}}{\alpha !\beta !}
D_k^{\beta} D_x^{\alpha} \sigma(k,x)
D_k^{\alpha} D_x^{\beta} \tau(k,x) + O(h^{\gamma})
\nonumber \\
&=& \sum_{j=0}^{\gamma} \frac{(i\pi h)^j}{j!}
[D_{k,\sigma} D_{x,\tau} - D_{k,\tau} D_{x,\sigma}]^j \sigma(k,x)
\tau(k,x)
+ O(h^{\gamma}),
\label{eq:series}
\end{eqnarray}
where the second subscripts $\sigma$ and $\tau$ of $D$ indicate the
symbol
to be acted upon.
The asymptotic expression for the symbol of the commutator of two
operators follows from \eqref{eq:series}. The even order terms
in the sum cancel out to produce
\begin{eqnarray}
[\sigma\sharp_h\tau - \tau\sharp_h\sigma](k,x)
&=& 2i \sum_{j=0}^{\gamma} \frac{(-1)^j(\pi h)^{2j+1}}{(2j+1)!}
[D_{k,\sigma} D_{x,\tau} - D_{k,\tau} D_{x,\sigma}]^{2j+1} \sigma(k,x)
\tau(k,x)
\nonumber \\
&+& O(h^{2\gamma+1}).
\label{eq:comseries}
\end{eqnarray}
The series expansion of the symbol of the commutator
\eqref{eq:comseries} allows to derive the Poisson bracket as the
classical limit of the symbol of the Heisenberg commutator of two
quantum
operators\begin{eqnarray}
\lim_{h\rightarrow 0} \frac{2\pi i}{h}[\sigma\sharp_h\tau-
\tau\sharp_h\sigma](k,x)
&=& \{\sigma(k,x),\tau(k,x)\} .
\label{eq:limit}
\end{eqnarray}
Since the commutative representations of the Heisenberg
group leave symbols of operators unchanged, i.e., $\int_{\Heisen{n}}
\F^{-1}[\sigma](g)\rho_{h=0}(g) dg= \int\int \F^{-1}[\sigma](p,q)
e^{2\pi
i(pk+qx)} dp dq = \sigma(k,x)$, we deduce that under the commutative
representations the \p-mechanical equation of motion (\ref{eq:eom})
reduces to the Hamilton equation
\begin{eqnarray}
\frac{\partial}{\partial t}A(k,x) &=&
\{H(k,x),A(k,x)\}.
\label{eq:hamilton}
\end{eqnarray}

In summary, the Heisenberg group contains the
exact quantum and classical descriptions of a system of particles
and provides the correspondence principle between the descriptions.
We refer the reader to Chapters~1 and~2 of reference~\cite{Folland89}
for further information on the subject.


\section{The Quantum-Classical Equation of Motion}\label{sec3}

We proceed to derive an equation of motion for a mixed
quantum-classical system.  In order to do this we look for
an abstract mathematical structure that has the same
role as the Heisenberg group in the case of the standard
quantization.  The desired structure can be constructed based on
the following observations.

First, we need two sets of observables $\{D,X\}$ and $\{D',X'\}$
corresponding to quantum and classical parts accordingly. We see no
reason to assume that an operator from the first set does not commute
with an operator from the second set.  We do assume that each set
has a Planck constant of its own. Then we let the Planck constant of
the second set approach zero and obtain the classical limit for the
second subsystem leaving the first one quantum. We know
that ``Planck constants'' arise as characters of the center.
Therefore, we need a
nilpotent Lie group with a two-dimensional center.

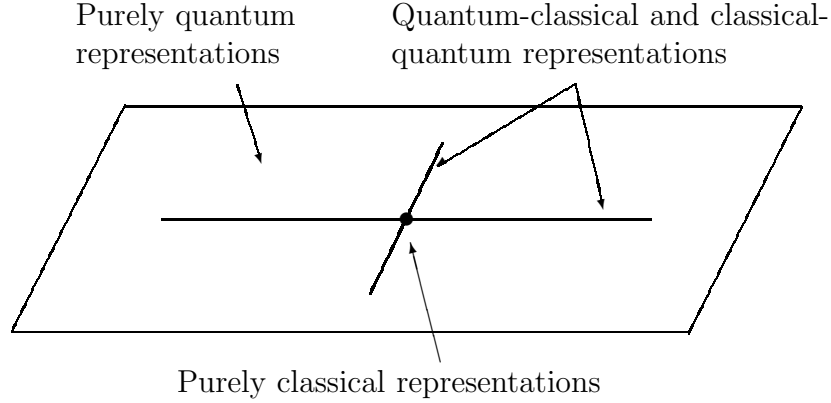
\begin{figure}[t]
\begin{center}
\unitlength 1.00mm
\linethickness{0.4pt}
\begin{picture}(115.00,50.00)
\multiput(10.00,10.00)(0.12,0.24){126}{\line(0,1){0.24}}
\put(25.00,40.00){\line(1,0){90.00}}
\multiput(115.00,40.00)(-0.12,-0.24){126}{\line(0,-1){0.24}}
\put(100.00,10.00){\line(-1,0){90.00}}
\thicklines%
\put(30.00,25.00){\line(1,0){65.00}}
\multiput(57.67,15.00)(0.12,0.25){81}{\line(0,1){0.25}}
%
\thinlines
\put(62.50,25.00){\circle*{1.75}}
\put(35.00,45.00){\makebox(0,0)[cb]
{\parbox{33\unitlength}{Purely quantum \\ representations}}}
\put(43.00,33.00){\vector(1,-3){0.2}}
\multiput(40.00,43.00)(0.12,-0.38){26}{\line(0,-1){0.38}}
\put(98.00,45.00){\makebox(0,0)[cb]
{\parbox{75\unitlength}{Quantum-classical and classical- \\ quantum
representations}}}
\put(88.50,27.00){\vector(1,-4){0.2}}
\multiput(85.00,43.00)(0.12,-0.53){30}{\line(0,-1){0.53}}
\put(66.60,32.20){\vector(-3,-2){0.2}}
\multiput(85.00,43.00)(-0.20,-0.12){92}{\line(-1,0){0.20}}
\put(67.00,5.00){\makebox(0,0)[ct]
{\parbox{70\unitlength}{Purely classical representations}}}
\put(67.00,6.00){\vector(-1,4){4}}
\end{picture}
\end{center}
\caption{Different types of descriptions generated by a step two
nilpotent Lie group with a two-dimensional center.}\label{fi:qc_b}
\end{figure}

This ``quantum-classical group'' is generated by two sets of variables
$\{hD,X\}$
and $\{h^{\prime}D^{\prime},X^{\prime}\}$ satisfying the
commutator relations
\begin{eqnarray}
[hD_j,X_k] = -ih\delta_{jk} I,\qquad
{[h^{\prime}D^{\prime}_{j'},X^{\prime}_{k'}]} = {-ih^{\prime}
\delta_{jk}
I^{\prime}}, \qquad 1\leq j,k \leq n;~~ 1\leq j',k' \leq
n'.\label{eq:Hcommutators}
\end{eqnarray}
Other commutators are zero. The group has a two-dimensional center
$\Z=\{sI+s^{\prime} I^{\prime}; s,s^{\prime}\in\R\}$.
Irreducible representations of a nilpotent Lie group are induced by
the characters of the center~\cite{Kirillov76}. For the 
quantum-classical group
the characters are
\begin{displaymath}
\mu: (z,z')\mapsto \exp(i(h z +h' z' )).
\end{displaymath}
\changeone{It is clear that for $hh'\neq 0$ the induced representation
coincides
with the irreducible representation of $\Heisen{n+n'}$ on
$\FSpace{L}{2}(\Space{R}{n+n'})$. This corresponds to \emph{purely
quantum} behavior of both sets of variables~(see
Definition~\ref{it:representation}). The trivial character $h=h'=0$
gives the family of
one-dimensional representations parameterized by  $\Space{R}{2(n+n')}$
and the \emph{purely classical} description.
These situations were studied in detail in the previous Section. A new
situation appears when $h\neq 0$ and $h'=0$, which
produce quantum behavior for the first set and classical behavior
for the second set. (The $h= 0$, $h' \neq 0$ case just permutes the
quantum and classical parts.)
Figure~\ref{fi:qc_b}
illustrates these facts. We find that
in the topology on the dual to the quantum-classical group
the quantum descriptions are dense in the quantum-classical and
classical descriptions, and the quantum-classical descriptions are
dense in the classical ones.}

Consider the quantum-classical case in more detail. The
quantum-classical representation is given by
\begin{eqnarray}
\rho_h(p,q,s,p^{\prime},q^{\prime},s')
&=& e^{2\pi i (s\cdot hI + p\cdot hD + q\cdot X +
p^{\prime}\cdot k^{\prime} + q^{\prime}\cdot x^{\prime})}.
\comment{&=& \exp({2\pi i (s\cdot h + p\cdot h \frac{\partial }{i\partial x}+
q\cdot xI + p^{\prime}\cdot k^{\prime} + q^{\prime}\cdot x^{\prime})}),}
\label{eq:rho_h0}
\end{eqnarray}
\changeone{where $k',x'\in \Space{R}{n'}$ and $h\in
\Space{R}{}\setminus \{0\} $.}
In this representation an element of the convolution
algebra on the quantum-classical group is identified with a
quantum-classical
operator acting of $L_2(\R^n)\otimes\R^{2n^{\prime}}$.
The operator can be computed
in terms of the Weyl transform of its symbol taken with respect
of the quantum (unprimed) coordinates
\begin{eqnarray}
\sigma(hD,X,k^{\prime},x^{\prime}) f(x) &=& \int
K_{\sigma}(x,y,k^{\prime},x^{\prime})
f(y) dy
\label{eq:Wtransforms_a}
\\ K_{\sigma}(x,y,k^{\prime},x^{\prime}) &=&
h^{-n} \int \sigma(k,\frac{x+y}{2},k^{\prime},x^{\prime}) e^{2\pi i(x-
y)k/h} dk.
\label{eq:Wtransforms}
\end{eqnarray}
The quantum-classical analog of the commutator is determined by
the limiting procedure
$h\neq 0$,
$h'\rightarrow 0$ used to derive
the Poisson bracket from the quantum commutator.
First we need to obtain the expression for the symbol of
the product of two operators.
We start with the expression analogous to \eqref{eq:product}, but
having two, rather than one set of coordinates. Focusing on the primed
variables, we carry out the transformations identical to those done
in deriving \eqref{eq:limit}:
\begin{eqnarray}
\sigma\sharp_h\tau(k,x,k^{\prime},x^{\prime})
&=& \bigg(\frac{2}{h}\bigg)^{2n} \int\!\int\!\int\!\int du dv d\eta
d\zeta
e^{4\pi i[(x-u)(k-\eta)-(x-v)(k-\zeta)]/h}
\nonumber \\
&\times& \bigg(\frac{2}{h'}\bigg)^{2n'} \int\!\int\!\int\!\int
du^{\prime} dv^{\prime} d\eta^{\prime} d\zeta^{\prime}
e^{4\pi i[(x'-u')(k'-\eta')-(x'-v')(k'-\zeta')]/h'}
\nonumber \\
&\times&\sigma(\zeta,u,\zeta^{\prime},u^{\prime})\tau(\eta,v,\eta^{\prime},v
^{\prime})
\nonumber \\
&=& \bigg(\frac{2}{h}\bigg)^{2n} \int\!\int\!\int\!\int du dv d\eta
d\zeta
e^{4\pi i[(x-u)(k-\eta)-(x-v)(k-\zeta)]/h}
\nonumber \\
&\times& \bigg(\frac{1}{h'}\bigg)^{2n'} \int\!\int du^{\prime}
dv^{\prime}
e^{4\pi i(v'-u')k'}
\F_3^{-1}[\sigma](\zeta,u,v^{\prime},x'+u^{\prime}h')
\nonumber \\
&\times&\F_3[\tau](\eta,v,u^{\prime},x'+v^{\prime}h')
\nonumber \\
&=& \bigg(\frac{2}{h}\bigg)^{2n} \int\!\int\!\int\!\int
du dv d\eta d\zeta e^{4\pi i[(x-u)(k-\eta)-(x-v)(k-\zeta)]/h}
\nonumber \\
&\times& \sum_{j=0}^{\gamma} \frac{(i\pi h')^j}{j!}
[D_{k^{\prime},\sigma(\zeta,u,k',x')}
D_{x^{\prime},\tau(\eta,v,k',x')}
- D_{k^{\prime},\tau(\zeta,u,k',x')}
D_{x^{\prime},\sigma(\eta,v,k',x')}]^j
\nonumber \\
&\times&[\sigma(\zeta,u,k',x')\tau(\eta,v,k',x')
] +
O(h'^{\gamma}).
\label{eq:products0}
\end{eqnarray}
At this point it is safe to drop the explicit dependence of the
symbols
on the primed variables, since the non-commuting nature of the symbols
is accounted for by the unprimed variables alone.
By taking the limit $h'\rightarrow 0$ we obtain the expression
for the symbol of the quantum-classical commutator
\begin{eqnarray}
[\sigma\sharp_h\tau - \tau\sharp_h\sigma](k,x)
&=& \bigg(\frac{2}{h}\bigg)^{2n} \int\!\int\!\int\!\int
du dv d\eta d\zeta e^{4\pi i[(x-u)(k-\eta)-(x-v)(k-\zeta)]/h}
\nonumber \\
&\times&\{\sigma(\zeta,u)\tau(\eta,v) - \tau(\zeta,u)\sigma(\eta,v)
\nonumber \\
&+&\lim_{h'\rightarrow 0} \frac{h}{h'}\sum_{j=0}^{\infty}
\frac{(-1)^j(\pi h')^{2j+1}}{(2j+1)!}
[D_{k^{\prime},\sigma(\zeta,u)} D_{x^{\prime},\tau(\eta,v)}
\nonumber \\
&-& D_{k^{\prime},\tau(\zeta,u)}
D_{x^{\prime},\sigma(\eta,v)}]^{(2j+1)}
[\sigma(\zeta,u)\tau(\eta,v) - \tau(\zeta,u)\sigma(\eta,v)]\}
\label{eq:seriess}
\end{eqnarray}
or
\begin{eqnarray}
[\sigma,\tau]_{\sharp_h}(k,x)
&=& \bigg(\frac{2}{h}\bigg)^{2n} \int\!\int\!\int\!\int
du dv d\eta d\zeta e^{4\pi i[(x-u)(k-\eta)-(x-v)(k-\zeta)]/h}
\nonumber \\
&\times&\bigg[\sigma(\zeta,u)\tau(\eta,v) -
\tau(\zeta,u)\sigma(\eta,v)
\nonumber \\
&+& \frac{ih}{2\pi}\bigg(\frac{\partial\sigma(\zeta,u)} {\partial k'}
\frac{\partial\tau(\eta,v)} {\partial x'}
- \frac{\partial\tau(\zeta,u)}{\partial k'}
\frac{\partial\sigma(\eta,v)} {\partial x'}\bigg)\bigg].
\label{eq:products}
\end{eqnarray}
\eqref{eq:products}
for the symbol of the commutator, together with
the rule for calculating operators from their symbols given by
Eqs.~(\ref{eq:Wtransforms_a}) and~(\ref{eq:Wtransforms}),
leads to the following equation of motion for a
mixed quantum-classical system\begin{eqnarray}
\frac{\partial}{\partial t}A(hD,X,k^{\prime},x^{\prime})
&=& \frac{2\pi i}{h} [H,A]_{\sharp_h} (hD,X,k^{\prime},x^{\prime}).
\label{eq:eoms}
\end{eqnarray}
This formula determines the evolution of operator $A$, which depends
on quantum and classical position and momentum variables and
acts on $L^2(\R^n)\otimes\R^{2n}$. The kernel of the
operator with respect to the $L^2(\R^n)$ subspace is given by
\eqref{eq:Wtransforms}.

The quantum-classical equation of motion (\ref{eq:eoms}) exhibits many
desired features.
If $A$ and $H$ depend solely on quantum or classical variables,
\eqref{eq:eoms} reduces to the corresponding purely quantum and purely
classical equations, Eqs.~(\ref{eq:heisenberg}) and
(\ref{eq:hamilton}) respectively.
Since the quantum-classical bracket of the right side of
\eqref{eq:eoms}
was obtained by selecting a representation for the Lie bracket of a
Lie group, it has all the properties of Lie brackets,
in particular,
it is antisymmetric and satisfies the Jacobi identity. It is
antisymmetric
because its symbol is antisymmetric: the integrand of
\eqref{eq:products}
changes sign under the permutation $\sigma\leftrightarrow\tau$.
To prove that the quantum-classical bracket satisfies the Jacobi
identity
we consider the expression for the symbol of the product of three
operators
$\sigma(hD,X,k',x')\tau(hD,X,k',x')\phi(hD,X,k',x')$
By successive application of \eqref{eq:products0} we get
\begin{eqnarray}
\sigma\sharp_h\tau\sharp_h\phi(k,x)
&=& \bigg(\frac{2}{h}\bigg)^{2n} \int\!\int\!\int\!\int du_1 dv_1
d\eta_1
d\zeta_1
e^{4\pi i[(x-u_1)(k-\eta_1)-(x-v_1)(k-\zeta_1)]/h}
\nonumber \\
&\times&\bigg(\frac{2}{h'}\bigg)^{2n} \int\!\int\!\int\!\int du dv
d\eta d\zeta
[\sigma(\zeta,u)\tau(\eta,v) - \tau(\zeta,u)\sigma(\eta,v)]
\nonumber \\
&\times&
\mbox{\boldmath $\{$}
[e^{4\pi i[(\zeta_1-u)(u_1-\eta)-(\zeta_1-v)(u_1-\zeta)]/h}
\phi(\eta_1,v_1)
\nonumber \\
&-& e ^{4\pi i[(\eta_1-u)(v_1-\eta)-(\eta_1-v)(v_1-\zeta)]/h}
\phi(\zeta_1,u_1)]
\nonumber \\
&+& i\pi h' \{ [D_{k'}\sigma(\zeta,u)\tau(\eta,v)
+ \sigma(\zeta,u)D_{k'}\tau(\eta,v)
\nonumber \\
&-& D_{k'}\tau(\zeta,u)\sigma(\eta,v)
- \tau(\zeta,u)D_{k'}\sigma(\eta,v)]
\nonumber \\
&\times& e^{4\pi i[(\zeta_1-u)(u_1-\eta)-(\zeta_1-v)(u_1-\zeta)]/h}
D_{x'}\phi(\eta_1,v_1)
\nonumber \\
&-& e^{4\pi i[(\eta_1-u)(v_1-\eta)-(\eta_1-v)(v_1-\zeta)]/h}
D_{k'}\phi(\zeta_1,u_1)
\nonumber \\
&\times&[D_{x'}\sigma(\zeta,u)\tau(\eta,v)
+ \sigma(\zeta,u)D_{x'}\tau(\eta,v)
\nonumber \\
&-& D_{x'}\tau(\zeta,u)\sigma(\eta,v)
- \tau(\zeta,u)D_{x'}\sigma(\eta,v)]\}
+ O(h'^2)
\mbox{\boldmath $\}$}.
\label{eq:jacobi1}
\end{eqnarray}
After taking the limit $\lim_{h'\rightarrow 0} \frac{h}{h'}$ of
the first and higher order terms in $h'$ and performing the
integrations,
this cumbersome expression can be rewritten in a more compact
symbolic form as
\begin{eqnarray}
\sigma\sharp\tau\sharp\phi &=& \sigma\tau\phi-\phi\tau\sigma
+ i(\sigma_{k'}\tau\phi_{x'} + \sigma\tau_{k'}\phi_{x'}
- \tau_{k'}\sigma\phi_{x'} - \tau\sigma_{k'}\phi_{x'}
\nonumber \\
&&-\phi_{k'}\sigma_{x'}\tau - \phi_{k'}\sigma\tau_{x'}
+ \phi_{k'}\tau_{x'}\sigma + \phi_{k'}\tau\sigma_{x'}),
\label{eq:jacobi2}
\end{eqnarray}
where the subscripts $k'$ and $x'$ indicate differentiation with
respect to these variables, and the ordering of the symbols
is to be kept track of. Given \eqref{eq:jacobi2} it is
straightforward to check that the Jacobi identity holds for the
symbols of
operators:
[$[\sigma,\tau]_{\sharp_h},\phi]_{\sharp_h}(k,x,k',x')$, and,
therefore, for the quantum-classical bracket.

\vspace{0.2in}

We used the Weyl correspondence principle to link quantum and
classical mechanics
and to derive the quantum-classical equation of motion.
It is well known, however, that the Weyl correspondence
in not unique in mapping phase space functions to Hilbert space
operators.
In fact, there exist arbitrary many such mappings, differing in the
order
assigned to products of position and momentum operators. The Weyl
rule leads to symmetrized products. For example, it maps $kx$ to
$\frac{1}{2}(hD\cdot X + X\cdot hD)$.  Another variant of the
correspondence
principle, widely used in the mathematics community because of its
simpler
form, is due to Kohn and Nirenberg. It keeps momentum operators on the
right,
mapping $kx$ to $X\cdot hD$.  It is straightforward to obtain a mixed
quantum-classical equation of motion using the Kohn-Nirenberg
calculus.
One starts with $\rho_h = e^{2\pi iq\cdot X} e^{2\pi ip\cdot hD}$
instead of $\rho_h = e^{2\pi i (p\cdot hD + q\cdot X)}$
and follows the same steps. The result is
\begin{eqnarray}
\frac{\partial}{\partial t}A(hD,X,k^{\prime},x^{\prime})
&=& \frac{2\pi i}{h} [H,A]_{\sharp_h}^{KN}
(hD,X,k^{\prime},x^{\prime})
\label{eq:eomkn}
\end{eqnarray}
with the correspondence rule
\begin{eqnarray}
\sigma(hD,X,k',x')^{KN} f(x) &=& \int K_{\sigma}^{KN}(x,y,k',x') f(y)
dy,
\nonumber \\
K_{\sigma}^{KN}(x,y,k',x') &=& h^{-n} \int \sigma(k,x,k',x') e^{2\pi
i(x-y)k/h} dk
\label{eq:kn}
\end{eqnarray}
and the following formula for the quantum-classical bracket of two
symbols
\begin{eqnarray}
[\sigma,\tau]_{\sharp_h}^{KN}(k,x,k',x')
&=& \bigg(\frac{2}{h}\bigg)^{2n} \int\!\int du dv  e^{4\pi i(x-u)(v-
k)/h}
\bigg\{\sigma(v,x)\tau(k,u) - \tau(v,x)\sigma(k,u)
\nonumber \\
&+& \frac{ih}{2\pi} \bigg[\frac{\partial\sigma(v,x)} {\partial k'}
\frac{\partial\tau(k,u)} {\partial x'} -
\frac{\partial\tau(v,x)}{\partial k'}
\frac{\partial\sigma(k,u)} {\partial x'}\bigg]\bigg\}.
\label{eq:KNproduct}
\end{eqnarray}
These expressions are somewhat simpler than those obtained by
the Weyl correspondence.
Eqs.~(\ref{eq:Wtransforms_a},\ref{eq:Wtransforms})
and~(\ref{eq:products},\ref{eq:eoms})
are preferable, however, as they preserve the simplectic
invariance of the phase space variables and lead via the Wigner
transform from the density matrix to the quantum quasi-probability
function
that is closest to the classical probability
density~\cite{Tatarskii83}.


\section{Quantum-Classical Coupling for Harmonic Oscillators}
\label{se:2oscilators}

The standard quantization can also be obtained by
application of a projection to the classical
system under consideration, which leads to a simple
description of harmonic oscillators. We briefly summarize this topic
and apply it to the case of coupled quantum and classical oscillators.
Further information can be found in
\cite{Berezin74,Berezin75,BergCob87,Coburn90,Coburn94a,CobXia94,%
Guillemin84} and references therein.

Let $\FSpace{L}{2}(\Space{C}{n},d\mu_{n})$ be a space of
functions on \Space{C}{n} square-integrable
with respect to the Gaussian measure
\begin{displaymath}
d\mu_{n}(z)=\pi^{-n}e^{-z\cdot\overline{z}}dv(z),
\end{displaymath}
where $dv(z)=dxdy$ is the Euclidean volume measure on
$\Space{C}{n}=\Space{R}{2n}$. The
Segal-Bargmann~\cite{Bargmann61,Segal60} (or Fock) space
$\FSpace{F}{2}(\Space{C}{n})$ is the subspace of
$\FSpace{L}{2}(\Space{C}{n},d\mu_{n})$ consisting of all
entire
functions, i.e., functions $f(z)$ that satisfy
\begin{displaymath}
\frac{\partial f}{\partial \bar{z}_j}=0, \qquad 1\leq j \leq n.
\end{displaymath}
Denote
by $P_Q$ the orthogonal Bargmann projection
\cite{Bargmann61}  of
$\FSpace{L}{2}(\Space{C}{n},d\mu_{n})$ onto the Fock
space $\FSpace{F}{2}(\Space{C}{n})$. Then
\begin{equation}
k(q,p)\rightarrow T_{k(q+ip)}=P_Q k(q+ip)I
\end{equation}
defines the Berezin-Toeplitz (anti-Wick) quantization, which
maps a
function
$k(q,p)=k(q+ip)$ on $\Space{R}{2n}=\Space{C}{n}$ to the
Toeplitz operator $T_k$. There exists an identification between
Berezin and Weyl
quantizations~\cite{Berezin74,Coburn90,Guillemin84}. The
identification has an especially transparent form for the observables
depending only on $p$ or $q$.

The Berezin-Toeplitz quantization is related to the Heisenberg group
more intuitively than the representation of~\eqref{eq:rho_h}.
On a geometrical level, consider the group of Euclidean shifts $a:
z\mapsto z+a$ of
\Space{C}{n}.  To obtain unitary operators on
$\FSpace{L}{2}(\Space{C}{n},d\mu)$ the shifts are
multiplied by the weight function
\begin{equation}\label{eq:fock-shift}
a: f(z)\mapsto f(z+a)e^{-z\bar{a}-a\bar{a}/2}.
\end{equation}
This mapping determines~\cite{Howe80b} a unitary representation of
the $(2n+1)$-dimensional Heisenberg group acting on
$\FSpace{L}{2}(\Space{C}{n},d\mu)$. The mapping preserves the Fock
space
$\FSpace{F}{2}(\Space{C}{n})$, and hence all
operators of the~\eqref{eq:fock-shift} form commute with $P_Q$.
The operators are generated by infinitesimal displacements
\begin{displaymath}
i \sum_{k=1}^{n}\left( a_j'(\frac{\partial }{\partial z_j'}-
z_j'-
iz_j'')+
a_j''(\frac{\partial }{\partial z_j''}-z_j''+iz_j')\right),
\end{displaymath}
where $a=(a_1,\ldots,a_n),
z=(z_1,\ldots,z_n)\in\Space{C}{n}$ and
$a_j=(a_j',a_j''), z=(z_j',z_j'')\in\Space{R}{2}$. The generators
form a linear space with the basis
\begin{equation}\label{eq:fshift-frame}
A^{f\prime}_j=\frac{1}{i}\left(\frac{\partial }{\partial
z_j'}-z_j'-
iz_j''\right),\quad
A^{f\prime\prime}_j= \frac{1}{i}\left(\frac{\partial
}{\partial z_j''}-
z_j''+iz_j'\right).
\end{equation}
The basis vectors commute with the Bargmann projector $P_Q$.
The operators
\begin{equation}\label{eq:fconv-frame}
X^{f\prime}_j=\frac{1}{i}\left(\frac{\partial }{\partial
z_j'}- z_j'+iz_j''\right),\quad
X^{f\prime\prime}_j= \frac{1}{i}\left(\frac{\partial }{\partial
z_j''}-
z_j''-iz_j'\right)
\end{equation}
commute with the basis vectors,
and we anticipate that $P_Q$ produces a self-adjoint representation of
convolutions with respect to $X^{f\prime}_j$, $X^{f\prime\prime}_j$,
and unit
operators.  Indeed,
\begin{prop}\textup{\cite{Kisil94e}}\label{pr:bargmann} The Bargmann
projector $P_Q$ defines a representation of convolutions induced by
the Weyl-Heisenberg Lie algebra $\algebra{h}_n$ operating on
\Space{C}{n}
by \textup{Eqs.~(\ref{eq:fconv-frame})}. The kernel $b(t,\zeta), t\in
\Space{R}{},\ zeta\in\Space{C}{n}$ of the representation is given by
the
formula
\begin{displaymath}
\widehat{b}(t,\zeta)=2^{n+1/2} e^{-1} e^{-
(t^2+\zeta\bar{\zeta}/2)}.
\end{displaymath}
\end{prop}

We move on to apply the Bargmann projection technique to
the quantum-classical coupling of harmonic oscillators.
\begin{example}~\cite{BergCob87} In the Segal-Bargmann
representation the operators of creation and annihilation of the $j$th
state
of a particle are $a^+_j=z_jI$ and
$a^-_j=\partial/\partial z_j$, correspondingly.
Consider an $n$~degree of freedom harmonic
oscillator with the classical Hamilton function
\begin{displaymath}
H(q,p)=\frac{1}{2}\sum_{j=1}^{n}(q_j^2 + p_j^2).
\end{displaymath}
The corresponding quantum Hamiltonian is obtained by the Bargmann
projection
\begin{equation}\label{eq:euler}
T_{H(q,p)}= \frac{1}{2} P_Q \sum_{j=1}^{n}(q_j^2 +
p_j^2)I=\frac{1}{2} (nI+\sum_{j=1}^{n}z_j\frac{\partial
}{\partial z_j}).
\end{equation}
The right side of \eqref{eq:euler} is the celebrated Euler
operator. It generates the well known dynamical
group~\cite[Chap.~1, Eq.~(6.35)]{MTaylor86}
\begin{equation}\label{eq:oscil-dyn}
e^{itT_{H(p,q)}} f(z)= e^{int/2} f(e^{it} z), \qquad f(z)\in
\FSpace{F}{2},
\end{equation}
which induces rotation of the $\Space{C}{n}$ space.  The evolution
of the classical oscillator is also given by a rotation, that
of the phase space $\Space{R}{2n}$
\begin{equation}
z(t)=G_t z_0= e^{it}z_0, \qquad z(t)=p(t)+iq(t),\
z_0=p_0+iq_0.
\end{equation}
The projection $P_Q$ leads to the Segal-Bargmann representation,
providing a very straightforward correspondence
between quantum and classical mechanics of oscillators, in contrast to
the rather
complicated case of the Heisenberg
representation~\cite[Prop.~7.1 of Chap.~1]{MTaylor86}.

The powers of $z$ are the eigenfunctions $\phi_n(z)=z^n$
of the Hamiltonian~(\ref{eq:euler}), and the integers $n$ are the
eigenvalues.
Either pure or mixed, any initial state of the oscillator remains
unchanged
during the~\eqref{eq:oscil-dyn} evolutions and
no transitions are observed.
\end{example}

Now consider classical and quantum oscillators coupled
by a quadratic term
\begin{equation}\label{eq:h-couple}
H(p,q;p',q')=\frac{1}{2}(p'^2 + p^2 +x'^2+x^2+\alpha (x'-x)^2).
\end{equation}
Applying the canonical transformation (see~\cite[\S~23.D]{Arnold91})
\begin{equation}
q'=\frac{q_1+q_2}{\sqrt[]{2}},\qquad
q=\frac{q_1-q_2}{\sqrt[]{2}}, \qquad
p'=\frac{p_1+p_2}{\sqrt[]{2}},\qquad
p=\frac{p_1-p_2}{\sqrt[]{2}}\label{eq:change},
\end{equation}
or, equivalently, introducing complex variables $z=q+ip$, $z'=q'+ip'$,
$z_1=q_1+ip_1$, $z_2=q_2+ip_2$
\begin{eqnarray}
z'=\frac{z_1+z_2}{\sqrt[]{2}},&& z=\frac{z_1-
z_2}{\sqrt[]{2}},\nonumber \\
z_1=\frac{z'+z}{\sqrt[]{2}},&& z_2=\frac{z'-z}{\sqrt[]{2}},
\end{eqnarray}
we get rid of the coupling term
\begin{equation}\label{eq:uncouple}
H(p_1,q_1;p_2,q_2)=\frac{1}{2}(p_1^2+p_2^2 + \omega_1^2 q_1^2 +
\omega_2^2 q_2^2),
\end{equation}
where $\omega_1=1$, $\omega_2=\sqrt[]{1+2\alpha }$.  The two uncoupled
oscillators
evolve independently
\begin{displaymath}
z_1(t)=e^{2i\omega_1t}z_1(t_0), \qquad
z_2(t)=e^{2i\omega_2t}z_2(t_0)
\end{displaymath}
The dynamics in the original coordinates, however, is not trivial.
The primed
and unprimed (quantum and classical) variables mix
\begin{eqnarray}
z'(t)&=&\frac{(e^{2i\omega_1 t}+e^{2i\omega_2 t}) z'(t_0)+
(e^{2i\omega_1 t}-e^{2i\omega_2 t}) z(t_0)}{2},\\
z(t)&=&\frac{(e^{2i\omega_1 t}-e^{2i\omega_2 t}) z'(t_0)+
(e^{2i\omega_1 t}+e^{2i\omega_2 t}) z(t_0)}{2}.
\end{eqnarray}

Suppose that the classical subsystem is initially localized at a point
$z'_0$ the phase space and the quantum subsystem is in its $n$-th
pure state $\phi(z',z;t_0)=\delta (z'_0-z')\otimes z^n$. The dynamics
of the combined system is given by
\begin{eqnarray}\label{eq:dyn1}
\phi(z',z;t)&=&\delta \left(z_0-\frac{(e^{2i\omega_1
t}+e^{2i\omega_2t})
z'+(e^{2i\omega_1 t}-e^{2i\omega_2 t}) z}{2}\right) \\
&&\otimes
\left( \frac{(e^{2i\omega_1 t}-e^{2i\omega_2 t}) z'+
(e^{2i\omega_1 t}+e^{2i\omega_2 t}) z}{2}\right)^n.\label{eq:dyn2}
\end{eqnarray}
\changeone{During the evolution the classical subsystem,
[\eqref{eq:dyn1}] is always
sharply supported, i.e. represented by the delta function, while the
quantum subsystem, [\eqref{eq:dyn2}] evolves into a mixed state.  The
binomial~(\ref{eq:dyn2}) contains all powers of $z$ less or equal to
$n$ ($z^k, k\leq n$). Therefore, there exists a non-zero probability
for the quantum subsystem to make a transition from the initial pure
state $z^n$ into a lower energy state $z^k$, $k<n$.
It is remarkable
that in this particular case the interaction with the classical
subsystem can only decrease the initial energy of the quantum one. The
overall dynamics is (quasi)-periodic with the recurrence time
determined by the frequencies $\omega_1$ and $\omega_2$.}

\section{Discussion and Conclusions}\label{sec4}

The quantum-classical equation of motion (\ref{eq:eoms}) can be
applied
in several ways depending on how one describes the quantum
and classical subsystems.  The classical subsystem can be treated on
the level of trajectories, in which case it is represented by
a point in the phase space $\{k'_i,x'_i\}$ evolving according to the
Hamilton
equations with the quantum-classical bracket of \eqref{eq:products}
regarded as
a modification of the Poisson bracket.
If at the same time the Heisenberg equation of motion with the
quantum-classical
bracket in place of the commutator is used to describe the evolution
of
quantum operators, the mean field approximation is recovered.
Namely, the quantum mechanical average of
\eqref{eq:eoms} with respect to the wave function $\Psi$ is given by
\begin{eqnarray}
\frac{\partial}{\partial t} \langle\Psi|A(k',x')|\Psi\rangle &=&
\frac{2\pi i}{h}
\langle\Psi|[H,A]_{\sharp_h}(k',x')|\Psi\rangle.
\label{eq:mf}
\end{eqnarray}
If $A$ is a purely quantum mechanical observable independent of
classical variables,
the derivatives $\partial A/\partial k'$ and $\partial A/\partial x'$
in
the quantum-classical bracket \eqref{eq:products} vanish, and we
obtain
\begin{eqnarray}
\frac{\partial}{\partial t} \langle\Psi|A|\Psi\rangle &=&
\frac{2\pi i}{h} \langle\Psi|[H(k',x'),A]|\Psi\rangle
\label{eq:mfq}
\end{eqnarray}
with the Hamiltonian $H$ parametrically depending on the classical
phase space variables $k'$ and $x'$.  Substituting the variables
in place of $A(k',x')$ in \eqref{eq:mf}, we recover the classical
equations of
motion, the classical Hamiltonian being the quantum mechanical average
of the
total Hamiltonian
\begin{eqnarray}
\frac{\partial k'}{\partial t} &=&
- \frac{\partial \langle\Psi|H(k',x')|\Psi\rangle}{\partial x'},
\nonumber \\
\frac{\partial x'}{\partial t} &=&
\frac{\partial \langle\Psi|H(k',x')|\Psi\rangle}{\partial k'}.
\label{eq:mfc}
\end{eqnarray}
Eqs.~(\ref{eq:mfq}) and~(\ref{eq:mfc}) constitute the traditional mean
field
approximation: classical variables are coupled to the expectation
values of
quantum observables.

The quantum-classical equation of motion can be looked upon in a
different way,
namely, as a Liouville-von~Neumann equation for a mixed
distribution $\rho(hD,X,k',x')$.
Selecting a quantum basis we get a set of coupled equations for
classical phase
space distribution functions $\rho_{ij}(k',x')$
corresponding to each pair of the quantum basis states $i$,~$j$
\begin{eqnarray}
\frac{\partial \rho_{ij}}{\partial t} &=& \frac{2\pi i}{h} \sum_k
\bigg[ H^*_{ik}\rho_{kj} - \rho^*_{ik}H_{kj} + \frac{ih}{2\pi} \bigg(
\frac{\partial H^*_{ik}}{\partial k'} \frac{\partial
\rho_{kj}}{\partial x'} -
\frac{\partial \rho^*_{ik}}{\partial k'} \frac{\partial
H_{kj}}{\partial x'} \bigg)
\bigg]
\label{eq:liouville_neumann}
\end{eqnarray}

In the purely quantum and classical limits the two derivatives
[Eqs.~(\ref{eq:mfq}),~(\ref{eq:mfc}) and
Eq.~(\ref{eq:liouville_neumann})]
of the quantum classical equation of motion [Eq.~(\ref{eq:eoms})] are
equivalent.
In the mixed case they are not, because of non-local correlations
in the classical subsystem induced by its interaction
with the quantum one.
Such correlations, inherent in the Liouville-von~Neumann equation,
do not appear on the level of individual trajectories.
In particular, if the equations~(\ref{eq:liouville_neumann}) are
integrated
with the initial conditions
$\rho_{11}(k',x')=\delta(k'-k_0')\delta(x'-x_0')$ and
$\rho_{ij}=0,~\forall~\{ij\}\ne\{11\}$,
at later times, in general, $\rho_{ij}\ne~0$ because of the couplings
$H_{ij}\ne~0$.
In other words, classical phase space distribution functions
associated
with different quantum states differ and mix.  Spreading of initially
localized
classical distributions is enhanced by their mutual mixing.
An initially localized
phase space distribution $\rho_{11}$ corresponding to the ground
quantum state
populates excited state distributions. They undergo diverging
evolutions, and
later, when $\rho_{11}$ gains an admixture of the excited state
distributions,
it necessarily becomes delocalized.  This, of course, can not happen
in the
mean field approach, where the classical subsystem is described by a
single
trajectory.  Non-local correlations are averaged out
in the mean field approximation.

If there is no coupling between quantum states, phase space
distributions
neither mix nor spread beyond the classical divergence.
Consider a purely adiabatic case, where the quantum basis
states are the instantaneous eigenstates of the quantum Hamiltonian.
In the absence of non-adiabatic
coupling \eqref{eq:liouville_neumann} splits into a set of uncoupled
equations for classical distribution functions corresponding to
individual adiabatic
quantum states.  An adiabatic evolution of every distribution function
can
be equivalently described by the classical trajectory mean field
approach
[Eqs.~(\ref{eq:mfq})--(\ref{eq:mfc})] with the wave function $\Psi$
being
the corresponding adiabatic eigenstate of the quantum Hamiltonian.

The Liouville-von~Neumann equation for the evolution of the mixed
distribution reduces to the coupled equations
of the mean field approximation if the quantum-classical function
is constrained to
\begin{eqnarray}
\rho &=& |\Psi\rangle\langle\Psi|\cdot\delta(k'-k'') \delta(x'-x''),
\label{eq:scmf}
\end{eqnarray}
where $k'$,$k''$ and $x'$,$x''$ are the $n$-dimensional classical
momentum and position
vectors.  Under this constraint the quantum part of the mixed function
always
remains a pure state, and the classical part is always represented by
a delta function.
Substituting expression~(\ref{eq:scmf}) for $\rho$ in place of $A$
in Eq.~(\ref{eq:eoms})
and integrating out the phase space variables ($\int dk'' \int dx''$)
we obtain the von~Neumann equation with the quantum Hamiltonian
depending on
classical coordinates $H(k',x')$.  Since the density matrix entering
the equation is constructed from a pure quantum state $\Psi$, the
von~Neumann
equation for the density matrix is isomorphic to the Schr\"{o}dinger
equation
and to Eq.~(\ref{eq:mfq}). To recover the mean field equations for the
classical
variables~(\ref{eq:mfc}) we substitute~(\ref{eq:scmf})
into~(\ref{eq:eoms}),
multiply both sides by either $k'$ or $x'$,
and integrate over quantum and classical coordinates.

The mean field approximation can not reproduce the non-local
correlations within
the classical
subsystem due to interaction with individual parts of the quantum
subsystem.
At the same time, such correlations naturally appear in the solutions
to
the Liouville-von~Neumann equation.  Unfortunately, the 
Liouville-von~Neumann equation
does not provide significant computational advantage over the pure
quantum
von-Neumann equation, since both deal with delocalized distributions.
On the other hand, propagation of an individual classical trajectory
via
the Hamilton equations of motion is far less demanding than
integration of
the Liouville equation.  The idea of running and then averaging over
a few representative trajectories
instead of propagating the total phase space distribution function is
fully
exploited in classical molecular dynamics simulations (see, for
instance,
reference~\cite{Allen&Tildesley}).
In order to account for the quantally induced non-local correlations
among the phase space variables, while retaining a trajectory
description for
the classical subsystem, we interpolate between the mean field
and Liouville-von~Neumann approaches by developing
a multiconfiguration version of the mean field method.
\comment{We propose to call it this way, since the underlying idea is
analogous to the idea of
the multiconfiguration self-consistent field approach discussed in
reference~\cite{Miller87}.}
We start with the quantum-classical distribution function
\begin{eqnarray}
\rho &=& \sum_i \sum_j \varrho_{ij} |\Psi_i\rangle\langle\Psi_j|\cdot
\delta(k'_{ij}-k') \delta(x'_{ij}-x'),
\label{eq:mcmf}
\end{eqnarray}
where the sums run over the number of independent configurations.
Taking
the wave functions to be orthonormal $\langle\Psi_i|\Psi_j\rangle =
\delta_{ij}$
we substitute~(\ref{eq:mcmf}) into Eq.~(\ref{eq:eoms}) and
integrate over the classical variables ($\int dk' \int dx'$) to obtain
the von~Neumann equation for the quantum density matrix
$\varrho = \sum_i \sum_j \varrho_{ij} |\Psi_i\rangle\langle\Psi_j|$
\begin{eqnarray}
\frac{\partial \varrho_{ij}}{\partial t} &=& \frac{2\pi i}{h} \sum_k
[H_{ik}^*(k'_{ik},x'_{ik}) \varrho_{ik} - \varrho_{kj}^*
H_{kj}(k'_{kj},x'_{kj})].
\label{eq:mcmfq}
\end{eqnarray}
To get the equations of motion for the classical variables $k_{ij}$,
$x_{ij}$
we substitute~(\ref{eq:mcmf}) into Eq.~(\ref{eq:eoms}), multiply both
sides
by the corresponding variable, and integrate over all degrees of
freedom.
We obtain equations of motion for the ``diagonal'' positions and
momenta as
\begin{eqnarray}
\frac{\partial k'_{ii}}{\partial t} &=&  - \sum_k
\frac{\partial H_{ki}(k'_{ki},x'_{ki})}{\partial x'_{ki}}, \nonumber
\\
\frac{\partial x'_{ii}}{\partial t} &=&
\sum_k \frac{\partial H^*_{ik}(k'_{ik},x'_{ik})}{\partial k'_{ik}} =
\sum_k \frac{\partial H_{ki}(k'_{ki},x'_{ki})}{\partial k'_{ki}}.
\label{eq:mcmfc}
\end{eqnarray}
The corresponding expressions for the ``non-diagonal'' variables are
more complicated
\begin{eqnarray}
\frac{\partial x'_{ij}}{\partial t} &=&
\sum_k  H^*_{ik}(k_{ik},x_{ik}) x'_{kj} - x'_{ik}
H_{kj}(k_{ik},x_{ik}) +
\frac{\partial H_{ik}(k_{ik},x_{ik})}{\partial k'_{ik}},
\label{eq:xij}
\end{eqnarray}
and similarly for momenta.
In order to keep the Hamiltonian matrix
$\parallel\! H_{ij}(k'_{ij},x'_{ij})\!\parallel$ hermitian
we require that the position and momentum
``matrices'' are symmetric: $k'_{ij}=k'_{ji}$, $x'_{ij}=x'_{ji}$.
With this constraint the ``non-diagonal'' evolutions simplify,
the first two terms in formula~(\ref{eq:xij}) disappear, and the
dynamics of
the ``non-diagonal'' phase space variables coincide with the average
dynamics of the ``diagonal'' variables
\begin{eqnarray}
\frac{\partial x'_{ij}}{\partial t} &\equiv&  \frac{1}{2} \bigg(
\frac{\partial x'_{ij}}{\partial t} + \frac{\partial x'_{ji}}{\partial
t} \bigg)
\nonumber \\ &=&
\frac{1}{2} \sum_k \bigg( H^*_{ik} x'_{kj} - x'_{ik} H_{kj} +
\frac{\partial H_{ik}}{\partial k'_{ik}} +
H^*_{jk} x'_{ki} - x'_{jk} H_{ki} +
\frac{\partial H_{jk}}{\partial k'_{jk}} \bigg)
\nonumber \\ &=&
\frac{1}{2} \sum_k \bigg( \frac{\partial H_{ik}}{\partial k'_{ik}} +
\frac{\partial H_{jk}}{\partial k'_{jk}} \bigg) =
\frac{1}{2} \bigg(
\frac{\partial x'_{ii}}{\partial t} + \frac{\partial x'_{jj}}{\partial
t} \bigg),
\nonumber \\
\frac{\partial k'_{ij}}{\partial t} &=&
\frac{1}{2} \bigg(
\frac{\partial k'_{ii}}{\partial t} + \frac{\partial k'_{jj}}{\partial
t} \bigg).
\label{eq:mcmfcc}
\end{eqnarray}
Apart from assigning the ``non-diagonal'' coordinates and momenta
unique values
a simple physical meaning: $x'_{ij}=(x'_{ii}+x'_{jj})/2$,
$k'_{ij}=(k'_{ii}+k'_{jj})/2$, Eqs.~(\ref{eq:mcmfcc})
reduce the number of independent classical trajectories in the
$n$-dimensional multiconfiguration mean field approximation from $n^2$
to $n$.

The idea of introducing multiple configurations is not new.
Apparently,
it originated in quantum chemistry as an improvement on the
self-consistent field
solution of the time independent Schr\"{o}dinger equation.
The time dependent fully quantum multiconfiguration
self-consistent field approach is discussed in
reference~\cite{Miller87}.
Equations of motion similar to our version of
the quantum-classical multiconfiguration
approximation were proposed by Diestler~\cite{Diestler83}.
The method was devised to account for non-local correlations among the
classical
degrees of freedom of a condensed phase solvent interacting with a
quantum solute.
\comment{by introducing multiple trajectories corresponding to
different electronic states.}
Our approach has a different derivation
and is less computationally demanding by a factor of $n$.
The number of configurations in the multiconfiguration mean-field
method does not have to be the same as the number of quantum
(adiabatic) basis states.
The former is usually less than the latter, and, in the case when the
classical
phase space can be separated into a union of several weakly connected
regions,
it should be determined by the number of such regions. For example, a
double well
system would require two configurations --- two classical
trajectories,
each originating in its own well.
\comment{Besides, Diestler formulated his approach for the wave functions
$|\Psi_i\rangle$
being adiabatic eigenstates of the Hamiltonian, while we do not place
any restrictions on the quantum basis. On the other hand, the
hemiquantal
equations of motion are easily generalized to general bases.}

The quantum-classical equation of motion is derived here via the
simplest extension
of the Heisenberg group of quantum and classical mechanics to the
group allowing for two sets of variables, such that the variables of
the first
set do not commute, while the variables of the second set do.
Obviously, there exist other groups satisfying this requirement.
It is possible, for instance, to consider a step three nilpotent Lie
group and the corresponding algebra decomposable as a vector space
into the three subspaces $V_0$, $V_1$, and $V_2$ having the following
properties:
The elements of $V_0$ commute with all elements and form the center.
Commutators of vectors from $V_1$ belong to $V_0$.  Commutators of
vectors from $V_2$ belong to $V_1$. By taking a representation
of this group that maps the center $V_0$ to zero, we would obtain
another model for a mixed quantum-classical system, where
vectors from $V_1$ would correspond to classical degrees of
freedom, since their commutators vanish, while vectors from $V_2$
would
describe quantum variables.  It is likely, though, that this scheme
will exhibit
properties atypical for quantum and classical mechanics,
since step three nilpotent Lie groups differ from step two groups
and the Heisenberg group in particular.
There exist, however, some advantages in dealing with
general nilpotent Lie groups.
\comment{For instance, an attempt to construct dynamics for a physical
system
with two distinct sets of quantum variables characterized by
\emph{different}
Planck constants by restricting attention to step two groups would
necessarily
lead to a model with both quantum and classical sets of variables.
This happens because all irreducible representations
of step two nilpotent Lie groups are products of the irreducible
representations of the Heisenberg group and the group of
transformations of the Euclidean space (see Chapter 6.2
of~\cite{MTaylor86}).
If the two Planck constants are equal, the Euclidean part of the
irreducible
representation vanishes, and a purely quantum picture is recovered.
If, however,
the two Planck constants differ, the Euclidean part is present, and,
therefore, the set
of classical variables is non-empty. Higher order nilpotent Lie groups
can help
to avoid this problem.}
\changeone{For example, the relativistic
quantization considered in reference~\cite{Kisil95b} is based on a
representation of the simplest step three nilpotent Lie group
(\emph{meta
Heisenberg group}~\cite{Folland94}) spanned by the
Schr\"odinger representation of the Heisenberg group and the operators
of
multiplication by functions.}
Application of the quantization rules to the appropriate
Lie algebras leads to quantum-classical constructions for
string theory, conformal field theory,
and Yang-Mills theories~\cite{Antonsen96}.

In summary, we presented a systematic approach to coupling quantum and
classical degrees of freedom based on a generalization of the unified
description
of quantum and classical mechanics in terms of
convolutions on the Heisenberg group.  Considering the simplest
extension of the Heisenberg group that allows for two distinct sets of
variables, we derived a quantum-classical equation of motion.
The quantum-classical bracket entering the equation is a Lie bracket
and, therefore, possesses the two most important properties common to
the
quantum commutator and the Poisson bracket: It is antisymmetric and
satisfies the Jacobi identity.  We explicitly constructed
the quantum-classical
equation of motion for coupled harmonic oscillators and discussed
approximations
to the equations applicable to more general cases.



\section{Acknowledgments}\label{sec7}

OVP thanks Arlen Anderson for providing copies
of unpublished works.  OVP would also like to thank Peter Rossky
and Bill Gardiner for useful discussions.
VVK is grateful to the Department of Mathematical Analysis,
University of Ghent and to the Fund of Scientific Research-Flanders,
Scientific Research Network ``Fundamental Methods and Technique in
Mathematics'' for support at the later stages of manuscript
preparation.


\newpage

\begin{thebibliography}{10}

\bibitem{Maddox95}
J. Maddox, Nature {\bf 373},  469  (1995).

\bibitem{Anderson95}
A. Anderson, Phys. Rev. Lett. {\bf 74},  621  (1995).

\bibitem{Boucher88}
W. Boucher and J. Traschen, Phys. Rev. D {\bf 37},  3522  (1988).

\bibitem{ChamConnes96b}
A. Chamseddine and A. Connes, The Spectral Actiion Principle (1996),
preprint
  \texttt{hep-th/9606001}.

\bibitem{Diestler83}
D.~J. Diestler, J. Chem. Phys. {\bf 78},  2240  (1983).

\bibitem{Aleksandrov81}
I.~V. Aleksandrov, Z. Naturforsch. {\bf 36A},  902  (1981).

\bibitem{Miller80}
H.~D. Meyer and W.~H. Miller, J. Chem. Phys. {\bf 72},  2272  (1980).

\bibitem{Tully76}
J.~C. Tully,  in {\em Dymanics of molecular collisions}, edited by
W.~H. Miller
  (Plenum, New York, 1976), p.\ 217.

\bibitem{Pechukas69a}
P. Pechukas, Phys. Rev. {\bf 181},  166  (1969).

\bibitem{Pechukas69b}
P. Pechukas, Phys. Rev. {\bf 181},  174  (1969).

\bibitem{Ehrenfest27}
P. Ehrenfest, Z. Physik {\bf 45},  455  (1927).

\bibitem{Hepp74}
K. Hepp, Comm. Math. Phys. {\bf 35},  265  (1974).

\bibitem{Mott31}
N.~F. Mott, Proc. Cambridge Phil. Soc. {\bf 27},  553  (1931).

\bibitem{Mittelman61}
M.~H. Mittelman, Phys. Rev. {\bf 122},  449  (1961).

\bibitem{Delos72a}
J.~B. Delos, W.~R. Thorson, and S.~K. Knudson, Phys. Rev. A {\bf 6},
709
  (1972).

\bibitem{Delos72b}
J.~B. Delos and W.~R. Thorson, Phys. Rev. A {\bf 6},  720  (1972).

\bibitem{Tully71}
J.~C. Tully and R.~K. Preston, J. Chem. Phys. {\bf 55},  562  (1971).

\bibitem{Tully90}
J.~C. Tully, J. Chem. Phys. {\bf 93},  1061  (1990).

\bibitem{OPrezhdo96b}
O.~V. Prezhdo and P.~J. Rossky,   (1996), submitted to \emph{J. Chem.
Phys.}

\bibitem{Kisil96a}
V.~V. Kisil, Journal of Natural Geometry {\bf 9},  1  (1996), e-print
archive
  \texttt{funct-an/9405002}.

\bibitem{Kisil94e}
V.~V. Kisil, Relative Convolutions. \irm. {Properties} and
Applications,
  Reporte Interno \#~162, Departamento de Matem\'aticas, CINVESTAV del
I.P.N.,
  Mexico City (1994), e-print archive \texttt{fuct-an/9410001}, to
appear in
  \emph{Adv. in Math.}

\bibitem{Dixmier69}
J. Dixmier, {\em Les {$C$}*-algebres et Leurs Representations}
  (Gauthier-Villars, Paris, {\noopsort{}}1964).

\bibitem{DaunHof68}
J. Dauns and K.~H. Hofmann, {\em Representation of Rings by Sections},
Vol.~83
  of {\em Memoirs Amer. Math. Soc.} (AMS, Providence, Rhode Island,
1968).

\bibitem{Hofmann72}
K.~H. Hofmann, Bull. Amer. Math. Soc. {\bf 78},  291  (1972).

\bibitem{Folland89}
G.~B. Folland, {\em Harmonic analysis in phase space} (Prinston Univ.
Press,
  Prinston, New Jersey, 1989).

\bibitem{Kirillov76}
A.~A. Kirillov, {\em Elements of the Theory of Representations}
  (Springer-Verlag, New York, 1976).

\bibitem{Tatarskii83}
V.~I. Tatarskii, Sov. Phys. Usp. {\bf 26},  311  (1983).

\bibitem{Berezin74}
F.~A. Berezin, Math. USSR-Izv. {\bf 8},  1109  (1974).

\bibitem{Berezin75}
F.~A. Berezin, Math. USSR-Izv. {\bf 9},  341  (1975).

\bibitem{BergCob87}
C.~A. Berger and L.~A. Coburn, Trans. Amer. Math. Soc. {\bf 301},  813
(1987).

\bibitem{Coburn90}
L.~A. Coburn, Proceedings of Symposia in Pure Mathematics {\bf 51},
97
  (1990).

\bibitem{Coburn94a}
L.~A. Coburn, {\em Algebraic Mettods in Operator Theory} (Birkh\"auser
Verlag,
  New York, 1994), pp.\ 101--108.

\bibitem{CobXia94}
L.~A. Coburn and J. Xia, Comm. Math. Phys. {\bf 168},  23  (1995).

\bibitem{Guillemin84}
V. Guillemin, Integral Equations Operator Theory {\bf 7},  145
(1984).

\bibitem{Bargmann61}
V. Bargmann, Comm. Pure Appl. Math. {\bf 3},  215  (1961).

\bibitem{Segal60}
I.~E. Segal, {\em Lectures at the Summer Seminar on Appl. Math.}
(Boulder,
  Colorado, 1960).

\bibitem{Howe80b}
R. Howe, J. Funct. Anal. {\bf 38},  188  (1980).

\bibitem{MTaylor86}
M.~E. Taylor, {\em Noncommutative Harmonic Analysis}, Vol.~22 of {\em
Math.
  Surv. and Monographs} (American Mathematical Society, Providence,
Rhode
  Island, {\noopsort{}}1986).

\bibitem{Arnold91}
V.~I. Arnold, {\em Mathematical Methods of Classic Mechanics}
(Springer-Verlag,
  Berlin, {\noopsort{}}1991).

\bibitem{Allen&Tildesley}
M.~P. Allen and D.~J. Tildesley, {\em Computer Simulations in Liquids}
(Oxford
  Univ. Press, Great Britain, 1990).

\bibitem{Miller87}
N. Makri and W.~H. Miller, J. Chem. Phys. {\bf 87},  3248  (1987).

\bibitem{Kisil95b}
V.~V. Kisil, Relativistic Quantization and Improved Equation for a
Free
  Relativistic Particle, e-print archive \texttt{quant-ph/9502022}
(1995),
  submitted to \emph{Phys. Essays}.

\bibitem{Folland94}
G.~B. Folland,  in {\em Fourier Analysis: Analytic and Geometric
Aspects},
  No.~157 in {\em Lect. Notes in Pure and Applied Mathematics}, edited
by W.~O.
  Bray, P.~S. Milojevi\'c, and \v{C}aslav V.~Stanojevi\'c (Marcel
Dekker, Inc.,
  New York, 1994), pp.\ 121--147.

\bibitem{Antonsen96}
F. Antonsen,   (1996), \ LANL e-print achive 
\texttt{quant-ph/9608042}.

\end{thebibliography}
\newcommand{\noopsort}[1]{} \newcommand{\printfirst}[2]{#1}
  \newcommand{\singleletter}[1]{#1} \newcommand{\switchargs}[2]{#2#1}
  \newcommand{\irm}{\textup{I}} \newcommand{\iirm}{\textup{II}}
  \newcommand{\vrm}{\textup{V}}

\end{document}
\\